\documentclass[11pt,tightenlines]{revtex4}
\usepackage{amsfonts}
\usepackage{amsmath}
\usepackage{txfonts}
\usepackage{graphicx}
\usepackage{amssymb}
\usepackage{color}
\usepackage{wrapfig,lipsum,booktabs}
\usepackage{floatrow}
\usepackage{epstopdf}
\usepackage{hyperref}
\usepackage[normalem]{ulem}

\newcommand{\beginsupplement}{
        \setcounter{table}{0}
        \renewcommand{\thetable}{S\arabic{table}}
        \setcounter{figure}{0}
        \renewcommand{\thefigure}{S\arabic{figure}}
     }
     
\begin{document}
\title{Gamma synchronization of the hippocampal spatial map---topological model}
\author{Edward Basso$^{1}$, Mamiko Arai$^{2}$ and Yuri Dabaghian$^{3,4*}$}
\affiliation{$^1$ Department of Physics, Rice University, 6100 Main St. Houston, TX 77005 US\\
$^2$Department of Mathematics, Tokyo Women's Christian University, 2-6-1 Zempukuji, Suginami-ku, Tokyo 167-8585, Japan\\
$^3$Jan and Dan Duncan Neurological Research Institute, Baylor College of Medicine, Houston, TX 77030, USA\\
$^4$Department of Computational and Applied Mathematics, Rice University, Houston, TX 77005, USA\\
$^{*}$e-mail: dabaghia@bcm.edu}
\vspace{17 mm}
\date{\today}
\vspace{17 mm}
\begin{abstract}
The mammalian hippocampus plays a principal role in producing a cognitive map of space---an internalized 
representation of the animal's environment. The neuronal mechanisms producing this map depend primarily 
on the temporal structure of the hippocampal neurons' spiking activity, which is modulated by the oscillatory 
extracellular electrical field potential. In this paper, we discuss the integrative effect of the gamma rhythm, one 
of the principal components of these oscillations, on the ability of the place cell ensembles to encode a spatial 
map. Using methods of algebraic topology and statistical physics, we demonstrate that gamma-modulation of 
neuronal activity generates a synchronized spiking of dynamical cell assemblies, which enables learning a spatial 
map at faster timescales.
\end{abstract}
\maketitle

\newpage

\section{Introduction}
\label{section:intro}

The mammalian hippocampus plays a key role in spatial cognition. Place cells, the principal hippocampal neurons, 
manifest remarkable spatial specificity of spiking activity. They fire only in select locations in the environment. 
These locations are known as their respective place fields \cite{Best}. As a result, the place cells' spike trains 
contain information about the animal's current location \cite{Jensen1}, its future \cite{Pfeiffer} and past \cite{Carr} 
navigation routes, both in the wakeful state and even in sleep \cite{Ji}. Moreover, damages to the hippocampal 
network produce impairments in spatial learning and difficulties in navigation planning \cite{Kim,Savage}. It is 
hence believed that the place cell ensembles encode an internalized ``map" that serves as a basis of animal's 
spatial awareness \cite{McNamara,OKeefe}. 

\textbf{Motivation}. An increasing amount of both direct \cite{Alvernhe,Wu,eLife,Diba,Poucet} and indirect
\cite{Gothard,Leutgeb,Wills,Touretzky,Cheng} experimental evidence suggests that this map is topological in nature, 
a rough-and-ready connectivity framework into which other brain regions integrate more detailed metrical information. 
A number of approaches have been deployed to understand the neuronal computations that could produce such a 
framework \cite{Chen1,Chen2,Curto,PLoS,Arai,Muller,Burgess,CRISP,Samsonovich,Schemas}. In particular, the approach 
proposed in \cite{Curto,PLoS,Arai} exploits the connection between the place fields covering an environment and the 
Alexandrov-\v{C}ech theorem, which points out the possibility of reconstructing the topology of a space $X$ from the 
pattern of overlaps between regions that cover $X$. The fact that the place fields produce a dense cover of the 
environment suggests that the pattern of overlaps between them contains the information required to represent the 
environment's topology, which may hold the key to the way in which the hippocampus encodes its topological map of s
pace. This observation is taken further by noticing that the domains where several place fields overlap are precisely 
the ones where the corresponding place cells cofire. In other words, the information about the overlap of place fields 
is represented via the place cell coactivity, which suggests that the Alexandrov-\v{C}ech construction can be carried 
out not only via geometric pattern of the place field overlaps, but also through the analysis of the place cell coactivities. 

\begin{figure} 
\includegraphics[scale=0.8]{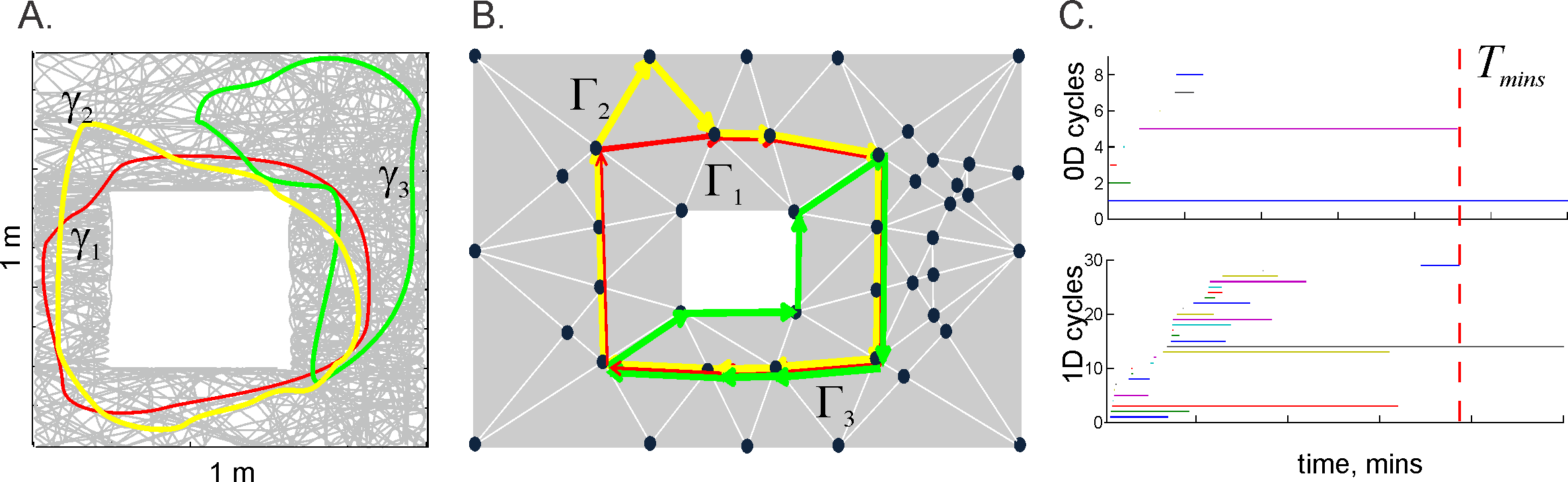}
\caption{\label{Figure1} \textbf{Paths in the environment are represented in a simplicial complex}.
(\textbf{A}) Two topologically equivalent paths in a physical environment, $\gamma_1$ and $\gamma_2$, encircle an obstacle 
(white square) that cannot traversed by the rat's trajectory (grey curve in the background) and are therefore non-contractible. 
The path $\gamma_3$ does not encircle the obstacle and therefore is contractible and topologically inequivalent to $\gamma_1$ 
and $\gamma_2$. (\textbf{B}) A schematic representation of the $2D$ skeleton of the coactivity complex $\mathcal{T}_{\sigma}$ 
(vertices shown as black dots, the $1D$ links as white lines and $2D$ facets as grey triangles) and of the simplicial paths $\Gamma_1$, 
$\Gamma_2$ and $\Gamma_3$, which represent the physical paths $\gamma_1$, $\gamma_2$ and $\gamma_3$. The topological 
equivalences and inequivalences between the simplicial paths ($\Gamma_1 \cong \Gamma_2$ and $\Gamma_1 \ncong \Gamma_3$, 
$\Gamma_2 \ncong \Gamma_3$) provide qualitative information about the physical paths, encoded via place cell coactivity. Since 
we are primarily concerned with representing topological properties of the navigational paths, in the following we discuss only the 
$2D$ skeleton of the coactivity complex. (\textbf{C}) Timelines of the topological loops encoded in the coactivity complex. As the 
animal begins to explore its environment, the coactivity complex contains many spurious topological loops most of which do not 
represent the physical obstacle. This ``topological noise" disappears after about five minutes, which marks the learning time, 
$T_{min}$---the moment when the correct topology of space (one $1D$ loop representing the obstacle and one $0D$ loop that 
informs us that the environment is connected) has emerged.} 
\end{figure}

\textbf{Topological model}. The details of the topological model of the hippocampal map are discussed in \cite{PLoS,Arai}, 
but in brief, the idea is to represent the combinations of coactive place cells ($c_1$, $c_2$, ..., $c_p$) as multi-dimensional 
polyhedra---the ``coactivity simplexes," $\sigma = [c_1, c_2, ..., c_p]$ (see Methods). Together, these coactivity simplexes 
form a simplicial ``coactivity complex" $\mathcal{T}_{\sigma}$. In his construction, the individual cell groups, just like 
simplexes, provide local information about the space; joined together into a simplicial complex, modeling a neuronal ensemble, 
they represent space as whole. Numerical simulations demonstrate that $\mathcal{T}_{\sigma}$ captures the topological 
structure of the environment and serves as a schematic representation of the hippocampal map \cite{PLoS,Schemas}. For 
example, the sequences of place cell combinations ignited along the paths traversed by the animal are represented in 
$\mathcal{T}_{\sigma}$ by chains of coactivity simplexes---the simplicial paths \cite{Novikov,Geom}. A non-contractible 
simplicial path may represent a navigational path $\gamma$ that encircles a physical obstacle, whereas topologically trivial 
simplicial paths correspond to contractible routes in the physical space (Figure~\ref{Figure1}A,B).

The complex $\mathcal{T}_{\sigma}$ begins to form as soon as the rat starts navigating. Every detected instance of place 
cell coactivity contributes a simplex to $\mathcal{T}_{\sigma}$. At the early stages of navigation, when only a few cells had 
time to produce spikes, the coactivity complex is small, fragmented, and contains many holes, most of which do not represent 
physical obstacles in the environment. Such holes, which may be viewed as transient ``gaps" in the cognitive map, tend to 
disappear as spatial learning continues. The minimal time, $T_{min}$, after which the topology of $\mathcal{T}_{\sigma}$ 
matches the topology of the environment, or more precisely, when the correct number of topological loops emerges, 
Figure~\ref{Figure1}C, can therefore be viewed as a theoretical estimate of the time required to learn the hippocampal map 
\cite{PLoS,Arai}. 

\textbf{Parameter dependence}. An important property of the model is that the structure of the coactivity complex $\mathcal{T}_{\sigma}$ 
and the time course of its formation during learning are sensitive to various parameters of the neuronal firing statistics, which allows 
studying the net effect produced by these parameters on spatial learning. For example, the oscillations of the extracellular electrical 
field potential, typically referred to as the local field potential (LFP), are known to modulate the place cells' activity at several timescales. 
First, each place cell tends to spike within a small range of the phases of the ``theta" component of the LFP ($\theta$, 4-12 Hz 
\cite{Buzsaki1}), which depends on the distance that the animal has traveled into the corresponding place field. As a rat moves 
through the place field, the preferred $\theta$-range of a place cell progressively decreases with each new $\theta$-cycle, a 
phenomenon known as the $\theta$-phase precession \cite{Huxter}. The preferred $\theta$-phases of different cells are additionally 
synchronized by the second major component of the LFP, the ``gamma" rhythm ($\gamma$, 30-80 Hz, \cite{Colgin1}). In fact, 
the period of the more rapid $\gamma$-rhythm, $T_{\gamma}$, is believed to define the range of the preferred phases within 
the slower $\theta$-rhythm; on average a $\theta$-period, $T_{\theta}$, contains about seven $\gamma$-cycles, 
$T_{\theta} \approx 7 T_{\gamma}$ (see \cite{Lisman1} and Figure~\ref{SFigure1}A).

Numerous experimental \cite{Shirvalkar,Tort,Lisman2,Nyhus,Duzel} and theoretical \cite{Lisman1,Koene,Jensen2,Zhang,Hasselmo1} 
studies demonstrate that both the $\theta$- and the $\gamma$-waves play key roles in spatial, working, and episodic memory 
functions. However, most theoretical analyses addressed the way in which the $\gamma$-synchronization affects the informational 
contents of spiking in small networks or in the individual cells. In contrast, the topological approach allows modeling cognitive map 
as a whole. For example, it was used in \cite{Arai} to demonstrate that $\theta$-precession makes otherwise poorly performing 
ensembles more capable of spatial learning. 

The present analysis applies the topological model to study the effect produced on spatial learning by the $\gamma$-waves and 
to demonstrate that $\gamma$-synchronization of the place cell spiking activity enables the encoding or retrieval of the large-scale 
spatial representations of the environment by integrating place cell coactivity at a synaptic timescale.

\section{The model}
\label{section:model}

Computational modeling of the $\theta$-phase precession is relatively straightforward: at the basic level, it amounts to imposing 
a particular relationship between a place cell's spiking probability, the phase of the $\theta$-wave and the distance that the animal 
has traveled into a corresponding place field \cite{Lengyel} (see Methods). However, the effects of the $\gamma$-rhythm are more 
diverse. Electrophysiological experiments suggest that there exist at least two types of place cells: the ``TroPyr" cells that spike at 
the trough of the fast $\gamma$-wave (50-80 Hz) and the ``RisPyr" cells that fire at the raising phase of slow $\gamma$-waves, 
overriding $\theta$-precession \cite{Senior,Osipova,Yamamoto}. Although our approach allows modeling both cell types (see Methods), 
in the following we will model only the TroPyr cells that exhibit more robust firing patterns and higher firing rates, and therefore may 
play a primary role in producing the cognitive map \cite{PLoS,Arai}.

\textbf{$\gamma$-modulation of spiking}. Physiologically, the $\gamma$-wave represents fast oscillations of the inhibitory postsynaptic 
potentials. As the amplitude of $\gamma$ drops at a certain location, the surrounding cells with high membrane potential spike 
\cite{Jia,Nikoli,Buzsaki2}. As a result, the preferred $\theta$-phase of several cells becomes synchronized with a $\gamma$-trough, 
which thereby gates the place cell coactivity. The literature refers to such groups of coactive place cells as ``dynamic cell assemblies" 
(see \cite{Harris1,Harris2,Buzsaki4}
and Figure~\ref{SFigure1}A).

Modeling $\gamma$-modulation therefore requires adjusting the times of the $\theta$-modulated spikes closer to the troughs of the 
$\gamma$-wave \cite{Colgin2}. Algorithmically, this task is similar to the task of distributing stochastic particles over the wells of a $1D$ 
potential energy field, which is solved based on the Maximum Entropy Principle \cite{Guiasu}. The probability that a particle lands at 
point $x$ in a potential $U(x)$ is $p \sim  e^{-\beta U(x)}$, where the parameter $\beta$ controls the spread of the locations around 
the minima of $U(x)$. In statistical physics, this parameter is interpreted as the inverse temperature. Higher values of $\beta$, 
meaning lower temperatures, imply that the particles are more confined to the bottoms of the wells (Figure~\ref{SFigure1}B).

\begin{figure} 
\includegraphics[scale=0.84]{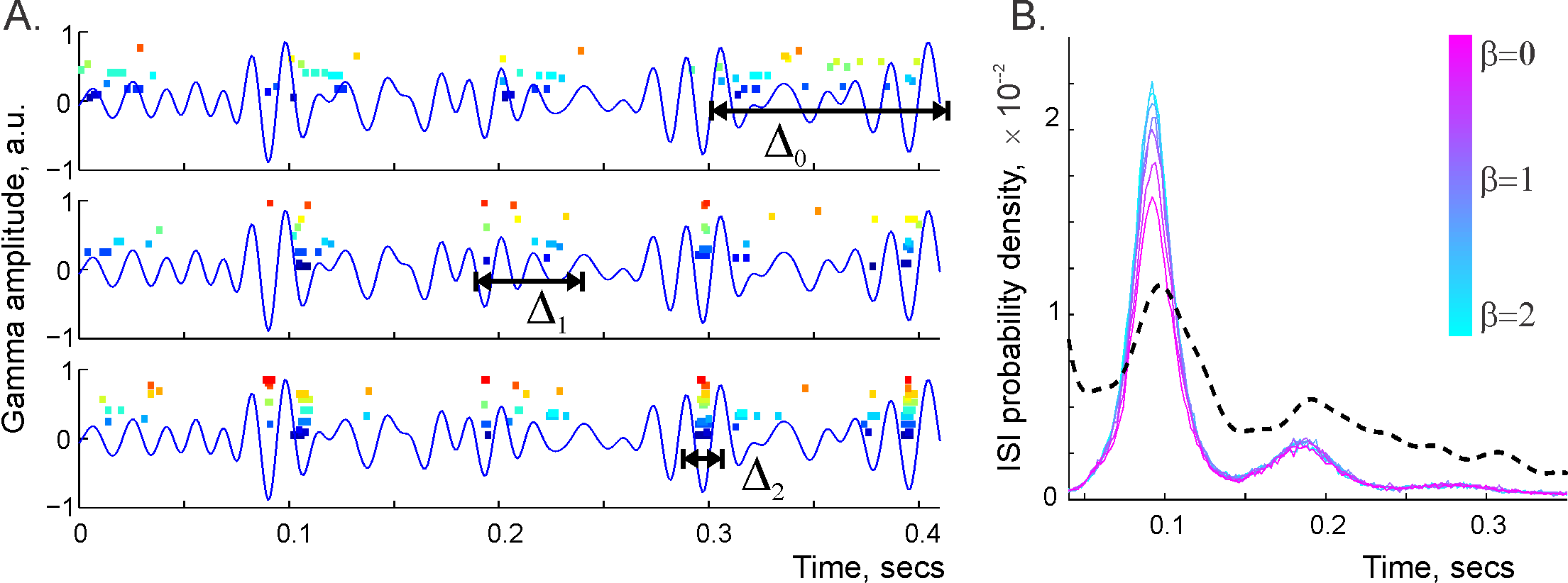}
\caption{\label{Figure2} \textbf{Gamma synchronization}. (\textbf{A}) Without coupling with the $\gamma$-wave ($\beta$ = 0, top panel) 
the simulated place cell spikes are diffusely scattered over the time axis. The temporal spread of the place cell coactivity is about a couple of 
$\theta$-periods, $\Delta_{0} \approx 1.5T_{\theta}$, $T_{\theta} \approx 125$ msec. At $\beta = 1$, the intervals of place cell coactivity 
concentrate near domains of high $\gamma$-amplitude, $\Delta_1 \approx 0.5T_{\theta}$. At $\beta =2$, the spikes accumulate near the 
$\gamma$-troughs, $\Delta_2 \approx T_{\gamma}$, thus producing dynamical cell assemblies (bottom panel). (\textbf{B}) The statistics 
of interspike intervals (ISI) for different $\beta$s. The black dashed line shows the distribution of the time intervals between deep 
$\gamma$-troughs (deeper than two standard deviations of $A(t)$ from the mean). As $\beta$ increases, the intervals between spikes are 
more controlled by the deep troughs (where the amplitude exceeds three standard deviations of $A(t)$ above the mean). Note, that the 
tendency of spikes produced by the same place cell to appear within the same $\gamma$-cycle can be viewed as a basic model of bursting 
\cite{Smith}.} 
\end{figure} 

Viewing the $\gamma$-amplitude, $A_{\gamma}(t)$, as an inhibitory potential extended over the time axis, we confined the place 
cells' firing to the troughs of the $\gamma$-wave by modulating their firing rates with the factor $e^{-\beta_{\gamma}A_{\gamma}(t)}$. 
Thus, the parameter $\beta_{\gamma}$ controls the temporal spread $\Delta_{\beta}$ of spikes produced by the dynamical cell assemblies. 
For small $\beta_{\gamma}$, the cell assemblies are ``hot," meaning their spikes are spread diffusely near the $\gamma$-troughs. 
For large $\beta_{\gamma}$ the assemblies are ``cold," their spikes ``freeze" at the $\gamma$-troughs (Figure~\ref{Figure2} and 
Figure~\ref{SFigure2}). In particular, the case in which the spike trains are uncorrelated with the $\gamma$-troughs corresponds 
to the limiting case of an ``infinitely hot" ($\beta_{\gamma} = 0$) hippocampus, modeled in \cite{Arai}. 

To our knowledge, the statistics of the temporal spreads of the spikes produced by the dynamical cell assemblies have not been studied. 
However, neurophysiological literature suggests that a typical spread is about one $\gamma$-period ($\Delta \approx T_{\gamma} 
\approx 20$ msec) \cite{Buzsaki3,Harris1,Harris2,Buzsaki4}, which implies that the effective temperature of the hippocampal cell 
assemblies is comparable to the mean $\gamma$-trough amplitude $1/\beta_{\gamma} \approx \bar A_{tr}$ (see Methods). In the 
following discussion, it will be convenient to scale the amplitude of the $\gamma$-wave with its standard deviation from the mean, 
$\sigma_{\gamma}$, $A_{\gamma}(t)\to A(t) = A_{\gamma}(t)/\sigma_{\gamma}$. In turn, this entails the corresponding scaling 
of the inverse temperature, yielding a parameter $\beta = \beta_{\gamma} \sigma_{\gamma}$ with the ``physiological" range 
between $0.5 \lesssim \beta \lesssim 2$.

\textbf{Reading out place cell coactivities}. The spiking signals produced by the place cells are transmitted to a population of readout 
neurons downstream from the hippocampus. In the reader-centric approach to information processing in the hippocampal network 
\cite{Schemas,Buzsaki4}, these neurons play a defining role. An assembly $\sigma$ is viewed not simply as an arbitrary combination 
of coactive place cells but as a functionally interconnected cell group that jointly triggers a spiking response from a certain readout 
neuron $n_{\sigma}$. In turn, the readout neuron $n_{\sigma}$ spikes upon receiving a sufficient amount of timed EPSP inputs over 
a certain period $w_{\sigma}$, the ``integration window" \cite{Magee1,Magee2,Spruston}. This is the only parameter that describes 
the readout neurons in the following discussion. Clearly, different readout neurons may integrate inputs over different time intervals. 
However, in order to simplify the present approach, we will describe the entire population of the readout neurons using a single parameter 
$w_{\sigma} = w$, viewed as the average time over which a typical readout neuron accumulates EPSP inputs \cite{Arai}.

\textbf{$\theta$ and $\gamma$ synchronicity}. In our previous study \cite{Arai}, we modeled assemblies of independently $\theta$-precessing 
place cells simply as combinations of neurons that happened to produce spikes within a certain $w$-period. The model predicted that the 
spatial maps are built reliably if the coactivity inputs are identified over the $\theta$-timescale ($T_{\theta} \lesssim w \lesssim 2T_{\theta}$). 
However, as $w$ shrinks, the chance of producing and detecting the coactivities within a $w$-period diminishes, and the topological map 
takes longer to form. For the intermediate range of values ($3T_{\gamma} \lesssim w \lesssim 3T_{\theta}$), the learning time is approximately 
inversely proportional to $w$, but as $w$ reduces to the $\gamma$-period, the pool of detected place cell coactivities often fails to capture the 
topological structure of the environment or requires a much longer time to produce it, exhibiting high variability of $T_{min}$ upon $w$. Moreover,
experimental studies have shown that the synchronicity of the place cell assemblies is best manifested precisely at the $\gamma$-timescale \cite{Harris1,Harris2,Buzsaki4}. This implies that the hippocampal network is capable of producing large-scale spatial maps based on the 
$\gamma$-timescale readouts. Therefore, we hypothesized that the failure of the previous ($\theta$-driven) topological model to do that may 
be due to poorer synchronization in the assemblies of independent neurons driven by common $\theta$-pacemaker, rather than to the physiological 
cell assemblies that are additionally synchronized through synaptic interactions \cite{Dragoi}. 

\begin{figure} 
\includegraphics[scale=0.78]{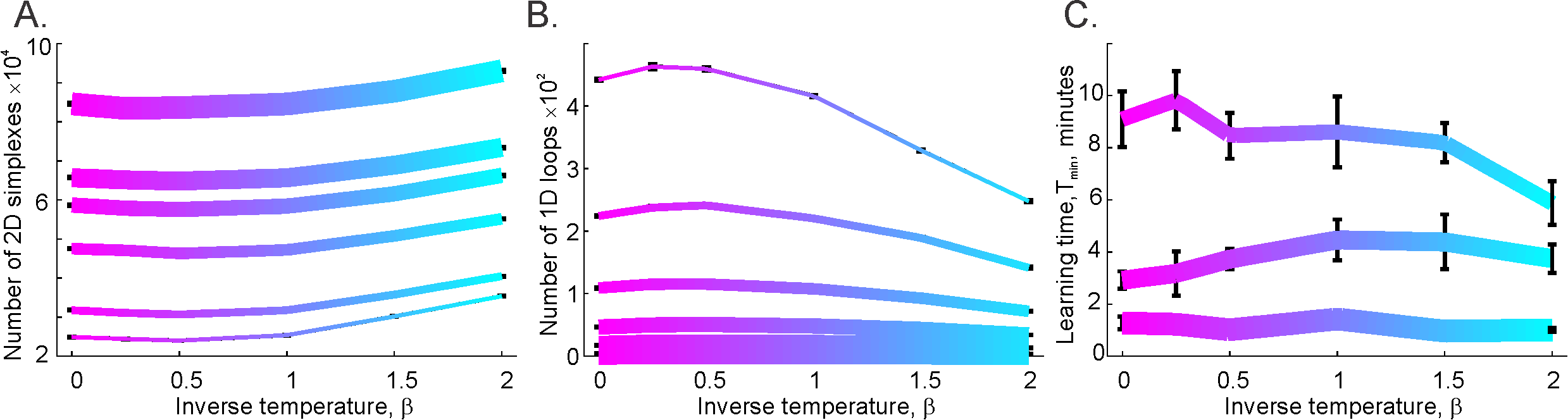}
\caption{\label{Figure3} \textbf{Influence of $\gamma$-modulation on spatial learning in a cell assembly network with coincidence detector 
readout neurons.}. There are two major parameters of the model: the mean width of the temporal window $w$ over which the postsynaptic 
readout neurons integrate spiking inputs from the place cell assemblies ($w_1=2 T_{\theta}$, $w_2=1.2 T_{\theta}$, $w_3=0.8 T_{\theta}$, 
$w_4=0.5 T_{\theta}$, $w_5=0.3 T_{\theta}$, $w_6=0.2 T_{\theta}$, represented by the thickness of the line), and the effective temperature 
$1/\beta$ which controls the clustering of place cells' spikes around the troughs of the $\gamma$-wave (Figure~\ref{Figure2}). Larger values 
of $\beta$ (indicated by the blue color of the colormap) correspond to tighter coupling between the place cell's spiking probability and the 
$\gamma$-amplitude (Figure~\ref{SFigure1}). (\textbf{A}) The number of $2D$ simplexes, $N_2$, in the coactivity complex $\mathcal{T}_{\sigma}$ 
as a function of $\beta$, for different $w$s. For large integration windows (thicker lines), coupling with $\gamma$-wave does not produce 
significant effect: the $N_2$ changes less along the $\beta$-axis. As $w$ decreases, the number of simplexes drops (the thinner lines lay 
below the thicker ones). However, the smaller is $w$, the more increase of the number of coactive place cell combination is produced by 
the cooling of the cell assemblies: for $w = T_{\gamma}$ (top curve) the number of $2D$ simplexes grows by $40\%$ as $\beta$ goes 
from $0$ to $2$. (\textbf{B}) Shrinking the integration window $w$ increases the total number of topological loops observed in $\mathcal{T}_{\sigma}$ 
during the course of learning, whereas cooling down the coactivity complex reduces this number. For example, the number of cold loops 
($\beta$ = 2) at $w = T_{\gamma}$ is about $50\%$ of the number hot loops ($\beta$ = 0). (\textbf{C}) The learning time $T_{min}$ grows 
as $w$ shrinks and tends to decrease as a function of $\beta$. Note however, even cold simplicial complexes fail to produce the correct 
topological maps for small $w$s.  In the particular map illustrated here (mean place field size $s = 24$ cm, mean firing rate $f = 20$ Hz, 
$N_c = 450$ cells), learning time diverges at $w\geq 0.5 T_{\theta}$.} 
\end{figure} 

In the present analysis, we use the effective temperature $1/\beta$ to describe phenomenologically these additional synchronization mechanisms. 
As illustrated in Figure~\ref{Figure2}, the parameter $\beta$ controls the temporal spread of the spiking activity in cell assemblies, $\Delta_{\beta}$,
independently from $w$ and allows transitioning from desynchronized cell assemblies to the cell assemblies that are tightly coupled with 
$\gamma$-troughs. The results shown in Figure~\ref{Figure2} also suggest that binding the coactivity of place cell assemblies within 
$\gamma$-periods ($\Delta_2 \approx T_{\gamma}$) should significantly reduce the time required by the downstream networks to detect place 
cell coactivity.  Thus, $\gamma$-synchronization may enable us to construct a reliable neuronal representation of space using the $\gamma$-timescale readout, $w \approx T_{\gamma}$, which would provide a direct demonstration of the importance of the $\gamma$-synchronization at the systemic level.

\section{Results}
\label{section:results}

To describe the effects of the $\gamma$-waves on the ability of place cells to encode spatial information, we built the coactivity complex
using $\gamma$-modulated spike trains for different $\beta$s and studied its topological properties for a set of $w$s, including the values 
for which the independently $\theta$-precessing place cells fail to produce correct topological maps. The results shown on Figure~\ref{Figure3} 
demonstrate that, at large integration windows ($w \geq T_{\theta}$, fat lines), tightening the cell assemblies around the $\gamma$-troughs 
does not produce a significant effect on either the structure of $\mathcal{T}_{\sigma}$ or on the times required to learn the map $T_min$. 
This outcome is easy to explain: if the readout neurons accumulate EPSPs at the $\theta$-timescale, i.e., over hundreds of milliseconds, the 
temporal arrangement of the spikes at the $\gamma$-timescale does not change the combinations of coactive place cell detected downstream. 
In other words, no matter how the $\gamma$-tuned spikes are spread inside a $\theta$-wide window $w$, the coactivity simplexes, and hence 
the coactivity complex, remain the same, yielding the same topological information after the same learning period. As $w$ decreases, the 
temporal spread of the poorly synchronized, ``hot" place cell assemblies begins to exceed $w$. As a result, only a fraction of the coactive 
cells can be detected downstream, which leads to a decrease of the number of simplexes in $\mathcal{T}_{\sigma}$ and to a proliferation 
of spurious topological loops during the learning period. Moreover, many of these loops persist indefinitely, preventing the appearance of the 
correct topological information even at the intermediate values of $w$ (Figure ~\ref{Figure3}C). 

In contrast, the behavior of the ``cold" cell assemblies ($\beta>1$, the blue ends of the graphs) is different. First, the number of $2D$ 
simplexes raises with cooling, which reflects the fact that the size of the cell assemblies increases with increasing $\beta$ (Figure ~\ref{Figure3}A). 
Second, colder coactivity complexes $\mathcal{T}_{\sigma}$ yield fewer, faster contracting spurious loops (Figure ~\ref{Figure3}B and Figure~\ref{SFigure3}). Third, the learning times drop significantly: for $\beta$ = 2, the $T_{min}$ computed for $w = 0.5 T_{\theta}$ 
reduce by about $50\%$ compared to the desynchronized, $\beta$ = 0 case, which indicates that $\gamma$-synchronization allows building 
a topological map based on the coactivity information transmitted to the downstream networks at times shorter than one $\gamma$-cycle
(Figure~\ref{Figure3}C and Figure ~\ref{SFigure4}). 

Nevertheless, the results shown on Figure ~\ref{Figure3} typically do not extend to the $\gamma$-timescale of $w$. The inputs collected 
from the cell assemblies which cooled to the physiological range of $\beta$s ($0.5 \lesssim \beta \lesssim 2$) at $w < 0.3 T_{\theta}$ often 
failed to produce an accurate map of the environment. This suggests that producing a correct neuronal map of space within a biologically 
plausible learning time using $w \approx T_{\gamma}$ requires further cooling of $\mathcal{T}_{\sigma}$ (by increasing $\beta$ indefinitely, 
the cell assemblies can be made as tight as desired). Thus, in order to keep the parameter $\beta$ within the physiological range, we have 
deployed an alternative approach.

\begin{figure} 
\includegraphics[scale=0.78]{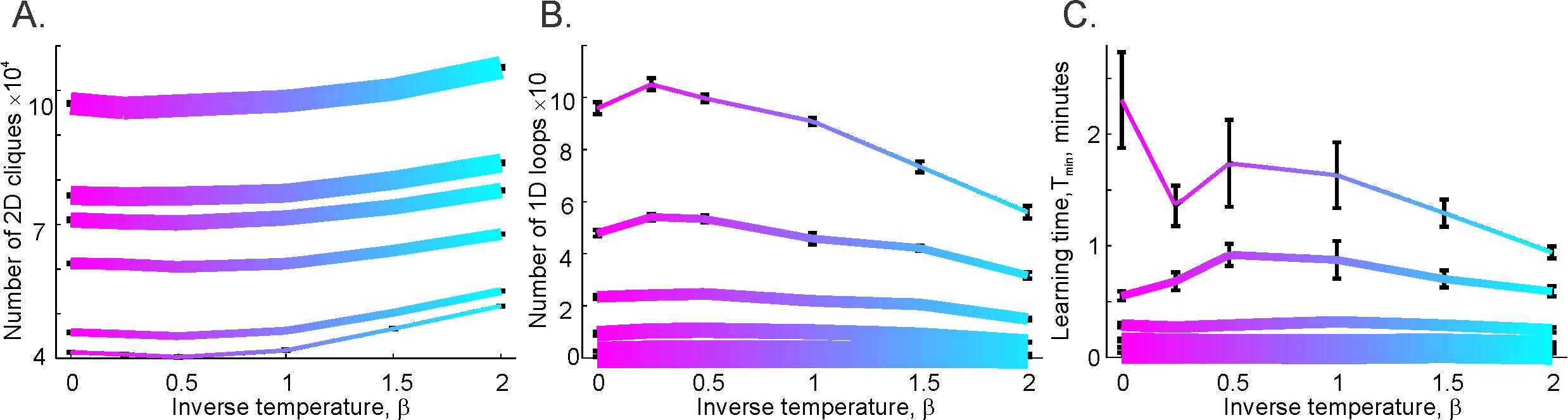}
\caption{\label{Figure4} \textbf{Influence of the $\gamma$-modulation on spatial learning in cell assembly network with input integrator 
readout neurons: clique complexes}. (\textbf{A}) The dependence of the number of triple connections in the clique coactivity complex 
$\mathcal{T}_{\varsigma}$ is similar to the dependence of number of $2D$ simplexes in the simplicial coactivity complex. As the integration 
window decreases (same range of $w$s as on Figure~\ref{Figure3}), the number of triple connections drops. Cooling down the assemblies 
does not produce significant effect at large integration windows, but increases the number of triple connections for small $w$s (by about $25\%$ 
for $w \approx T_{\gamma}$). 
(\textbf{B}) The total number of topological loops observed in the clique coactivity complex $\mathcal{T}_{\varsigma}$ is reduced with cooling 
for small $w$s, similarly to the case of the \v{C}ech coactivity complex. At the $\gamma$-timescale, $w \approx T_{\gamma}$, the tendency
of the shrinking $w$s to produce large numbers of topological loops in the clique coactivity complex is nearly compensated by cooling down
$\mathcal{T}_{\varsigma}$: the number of cold loops ($\beta = 2$) in $\mathcal{T}_{\varsigma}$ is about $50\%$ of the number hot loops 
($\beta = 0$). Note that despite similar qualitative behaviors, the scales of $N^{\sigma}_{l_1}$, and $ N^{\varsigma}_{l_1}$ are different: 
the clique complex produces fewer spurious loops than the simplicial complex. (\textbf{C}) The learning times grow as a function of $w$; 
however, for the clique complex they remain finite even for $w \approx T_{\gamma}$. Thus, the cooler the ensemble of cell assemblies, the 
faster it learns, especially for small $w$.} 
\end{figure} 

\textbf{Clique coactivity complexes}. In the above discussion, the central construction of the model, which is the \v{C}ech coactivity complex 
$\mathcal{T}_{\sigma}$, was introduced as a schematic representation of the place field map \cite{Schemas}. However, as shown in 
\cite{Comb,Kentaro}, a coactivity complex can be built not only by detecting higher order coactivity events that directly mark the locations were 
several place field overlap, but also by integrating the information provided by the lower order place cell coactivities. Physiologically, the latter 
option corresponds to the cell assembly network in which the readout neurons integrate lower order coactivity inputs over a working memory 
timescale, rather than merely react to coactivities as all-or-none coactivity detectors \cite{Konig,Ratte}. 

To model a network of cell assemblies driving a population of input-integrator readout neurons, we used the following approach. First, we detected 
the lowest order, pairwise place cell coactivities and used them to build a connectivity graph $G$ (see \cite{Comb} and Figure~\ref{SFigure5}). 
Then the maximal cliques of $G$ (see Methods) were identified with the maximal simplexes of a new ``clique" coactivity complex $\mathcal{T}_{\varsigma}$. 
A key property of this algorithm is that the connections constituting a clique or a simplex do not have to be detected at once---instead, they can be 
accumulated over an extended period. For physiological accuracy, we restrict this period to 10 mins or less, which results in a coactivity complex whose
simplexes, or a cell assembly network whose cell assemblies, emerge over working memory intervals.

Although the algorithm of constructing temporal \v{C}ech and the clique complexes seem quite different, the actual difference between these two 
coactivity complexes is not as significant. First, as shown in \cite{Schemas,Comb}, most simplexes of $\mathcal{T}_{\varsigma}$ correspond to the 
simplexes of $\mathcal{T}_{\sigma}$ and vice versa (i.e., the identities of the cell assemblies are largely the same, only the time course of their 
construction changes) and the topological structures of these complexes are quite close. Second, most pairwise connections within the cliques of 
$G$ are produced almost simultaneously while the rat traverses the region where several place fields overlap. In other words, most cliques appear 
at once, just as the simplexes do, and only a relatively small number of the maximal cliques are actually ``corrected" over time \cite{Kentaro}. 
Nevertheless, this effect does improve the overall performance of the clique coactivity complexes, which typically produce much smaller numbers of 
spurious topological loops. This corresponds to a shorter learning times $T_{min}$ than the \v{C}ech coactivity complexes.

Implementing the $\gamma$-synchronization mechanism in an integrator model yields the results illustrated in Figure~\ref{Figure4}. First, the 
structure of the graphs on Figures~\ref{Figure3}A and Figure~\ref{Figure4}A is qualitatively similar, though the pool of the third order cliques 
comes out to be slightly larger than the pool of $2D$ simplexes. This is because not every clique makes a simultaneous appearance as a simplex, 
but every simplex can be viewed as an instantly detected clique. The behaviors of the topological loops in $\mathcal{T}_{\varsigma}$ and in $\mathcal{T}_{\sigma}$, shown in Figures~\ref{Figure3}B and Figure~\ref{Figure4}B are similar as well: the $\gamma$-synchronization reduces 
the number of the cold spurious loops in both types of complexes (Figure~\ref{SFigure6}). Physiologically, this implies that $\gamma$-rhythm 
produces the same organizing effect on the activity of cell assembly network, whether the latter is based coincidence detector or on the input 
integrator readout neurons. 
However, it should be noted that, for all $\beta$s, the number of loops in $\mathcal{T}_{\varsigma}$ is smaller than in $\mathcal{T}_{\sigma}$ 
by an order of magnitude, which illustrates the efficiency of the input integrating readout neurons. Most importantly, the integrator complex $\mathcal{T}_{\varsigma}$ produces finite learning times at the $\gamma$-timescale integration window, $w \approx T_{\gamma}$, which 
demonstrates that the hippocampal 
network can produce a spatial memory map by reading out $\gamma$-synchronized place cell coactivity at the $\gamma$-timescale and accumulating 
them over working memory timescale, and the model provides a simple phenomenological mechanism for this demonstration.

\section{Discussion}
\label{section:discuss}

The neuronal activity is synchronized across the hippocampal network, giving rise to rhythmic flows of synaptic currents. The resulting waves 
of the mean extracellular field define the timescales of the place cell (co)activity and hence control the ``parcellation" of the information flow 
which is received by downstream networks. In particular, the synchronization of the processes taking place at the synaptic timescale, such as 
the processes controlled by the membrane time constant, by the duration of receptor-mediated postsynaptic spike potentials, by the rate of 
spike-timing dependent plasticity and so forth (\cite{Johnston,Magee3,Bi}) is manifested at the network level as $\gamma$-frequency oscillations
\cite{Traub,Atallah,Bartos,Csicsvari,Whittington1}. Processes that involve slower forms of synaptic plasticity, including slow changing spiking 
thresholds \cite{Huerta,Monyer,Henze,Mickus}, synchronize at the timescales of $\theta$-frequencies. As a result, $\theta$-oscillations provide 
lower resolution temporal packaging of place cell coactivity \cite{Dragoi,Hasselmo2,Hasselmo3}, which allows integrating spiking inputs from 
several cell assemblies over one or more $\theta$-periods \cite{Ang,Maurer,Mizuseki}. 

The topological model based on independently $\theta$-precessing place cells provides a self-consistent description of the hippocampal network's 
function at the $\theta$-timescale, predicting, in particular, an optimal integration window for reading out the information within the $\theta$-range 
\cite{Arai}. However, as the integration window becomes smaller, the spatial map encoded by independently precessing place cells fails to 
represent spatial maps, indicating the importance of additional synchronization at the $\gamma$-timescale and suggesting that further refinement 
of the model is required. 

The phenomenological model proposed above is based on the assumption that the $\gamma$-rhythm controls not only the probability of the cell 
assemblies' spiking but also defines the temporal spread of the spikes produced by the cell assemblies around the troughs of the $\gamma$-wave. 
As a result, the model predicts that if the preferred $\theta$-phases synchronize with the $\gamma$-troughs, the topological map of space can be 
robustly captured by integrating the place cell coactivity at the $\gamma$-timescale. Thus, $\gamma$-synchronization of spiking activity is crucial 
for encoding and reading out the large scale information by acquiring the information from the cell assembly inputs arriving in ``$\gamma$-packets" \cite{Buzsaki4}.

Second, the model can explain why a suppression of the $\gamma$-wave amplitude, induced by the changes in network's synaptic physiology 
\cite{Hormuzdi,Buhl,McHugh,Frisch,Cho}, or produced by psychoactive drugs \cite{Whittington2} such as cocaine \cite{McCracken,Dilgen}, 
or arising due to neurodegeneration or aging \cite{Vreugdenhil,Lu}, usually correlates with impairments in learning. Lastly, these results may also 
explain the well-known correlation between successful learning and retrieval of the learned information with the increase of the $\gamma$-amplitude 
in raised attention states \cite{Howard,Vugt,Moretti,Lundqvist,Trimper}.

\section{Acknowledgments}
\label{section:ackn}

We thank Robert Phenix for editing the manuscript. The work was supported in part by the NSF 1422438 grant (E.B. and Y.D.) Houston Bioinformatics Endowment Fund the W. M. Keck Foundation grant for pioneering research (M.A. and Y.D.)

\section{Methods}
\label{section:methods}

\textbf{Glossary}. An \emph{abstract simplex} of order $d$, $\sigma^{d}$, is a set of $(d +1)$ elements, e.g., a set of $(d +1)$ active 
cells. Note that the subsets of the set $\sigma^{d}$ form subsimplexes of $\sigma^{d}$ and that a nonempty overlap of any two simplexes 
$\sigma^{d}_1$ and $\sigma^{d}_2$ is a subsimplex of both $\sigma^{d}_1$ and $\sigma^{d}_2$. A \emph{simplicial complex} 
$\Sigma_{\sigma}$ is a family of simplexes. The elements of a simplex $\sigma^{d}$ can be visualized as vertices of $d$-dimensional polytopes: 
$\sigma^{0}$ can be visualized as a point, $\sigma^1$ as the ends of a line segment, $\sigma^2$ as the vertices of a triangle, 
$\sigma^3$ as the vertices of a tetrahedron, etc. \cite{Aleksandrov}.
A \emph{clique} in a graph $G$ is a set of fully interconnected vertices (i.e., a complete graph). Combinatorically, cliques have the same 
key properties as the abstract simplexes: any subcollection of vertices in a clique is fully interconnected, and hence forms a subclique. 
A nonempty overlap of two cliques $\varsigma^{d}_1$ and $\varsigma_2^{d}$ is a subclique in both $\varsigma^{d}_1$ and $\varsigma^{d}_2$. Thus, cliques define 
abstract simplexes and hence the collection of cliques in a graph $G$ defines a \emph{clique simplicial complex} $\Sigma_{\varsigma}(G)$.

\textbf{Choice of the simulated environment}. In \cite{Arai} we showed that the time required to learn a large spatial environment is 
approximately equal to sum of times required to learn its parts. We therefore simulated a non-preferential exploratory behavior in a 
small planar environment $(1m \times 1m)$ shown in Figure~\ref{Figure1}A, which is similar to the ones used in electrophysiological 
experiments \cite{Mamad}.

The \textbf{Poisson spiking rate} of a place cell $c$ at a point $r(t) = (x(t), y(t))$ is given by
\begin{equation}
\lambda_c(r)=f_c e^{-\frac{(r-r_c)^2}{2s^2_c}}
\nonumber
\end{equation}
where $f_c$ is the maximal firing rate and $s_c$ defines the size of the place field centered at $r_c = (x_c, y_c)$. The set of $s_c$s 
and $f_c$s in an ensemble of $N$ place cells are lognormally distributed around a certain ensemble-mean firing rate, $f$ and a 
certain ensemble-mean place field size $s$, with the variances $\sigma_f = af$ and $\sigma_s = bs$ respectively, i.e., a place cell 
ensemble is described by a triple of parameters: $(s,f,N)$ \cite{PLoS}.

\textbf{$\theta$-phase precession}. As the rat moves over a distance $l(t)$ into the place field of a cell $c$, the preferred spiking phase is 
\begin{equation}
\varphi_{\theta,c}(t)\approx 2\pi (1-l(t)/L_c),
\nonumber
\end{equation}
where $L_c \sim  3s_c$ is the size of the place field \cite{Huxter,Buzsaki5}. To simulate the coupling between the firing rate and the 
$\theta$-phase, we modulated the original Gaussian firing rate by a $\theta$-factor $\Lambda_{\theta,c}(\varphi)$, giving 
\begin{equation}
\Lambda_{\theta,c}(\varphi)=e^{-\frac{(\varphi - \varphi_{\theta,c}(t))^2}{2\varepsilon_c^2}},
\nonumber
\end{equation}
using the $\theta$-component of the LFP recorded in wild type mice. The width, $\varepsilon$, of the Gaussian was defined in 
\cite{Arai} to be the ratio of the mean distance that rat travels during one $\theta$-cycle to the size of the place field, 
$\varepsilon = 2\pi v /L\omega_{\theta}$, where $v$ is the rat's speed and $\omega_{\theta}/2\pi$ is the frequency of the $\theta$-signal.

\textbf{$\gamma$-modulation}. To incorporate the $\gamma$-rhythm into our model, we extracted the 30-80 Hz frequency band 
from the same LFP signal so that all the existing correlations between $\theta$ and $\gamma$ waves are preserved, and shifted the
simulated place cells' spiking times towards the troughs of $\gamma$ amplitude by modulating their respective spiking rates with 
the additional Boltzmann factor \cite{Guiasu},
\begin{equation}
\Lambda_{\gamma}(t)\sim e^{-\beta_{\gamma}A_{\gamma}(t)},
\label{Boltz}
\end{equation}
where $A_{\gamma}(t)$ is the amplitude of the $\gamma$-wave and $1/\beta_{\gamma}$ is a formal parameter that plays the 
role of the effective temperature \cite{Jaynes} (Figure~\ref{Figure2}). Simulating the net firing rate as a product of all three factors
$$\lambda_{net} = \lambda_{c}(x,y)\Lambda_{\theta,c}(\varphi)\Lambda_{\gamma}(A_{\gamma})$$
preserves spatial selectivity of spiking and the $\theta$-precession (Figure~\ref{SFigure7}) and forces the preferred phases 
of the $\theta$-phase precession $\varphi_c$ into the $\gamma$-cycles, in accordance with the $\theta$-$\gamma$ theory 
\cite{Lisman1,Lisman2,Lisman3}.

\textbf{Temperature of the cell assemblies}. In a vicinity of the $i^{th}$ trough, the gamma signal has the form 
\begin{equation} 
A_{\gamma}(t) \approx A_{\gamma,0}-A_{\gamma,i } \cos(\omega_{i} t) \approx a_{\gamma,i }+A_{\gamma,i}\frac{\omega^2_i t^2}{2},
\label{exp}
\end{equation}
where the parameters $A_{\gamma,0}$, $A_{\gamma,i}$ and $\omega_i$ define the mean level of $A_{\gamma}$, its instantaneous 
amplitude, and the instantaneous frequency at the $i^{th}$ trough, and $a_{\gamma,i} = A_{\gamma,0}- A_{\gamma,i}$. Using the 
expansion (\ref{exp}) in (\ref{Boltz}) allows estimating the spread $\Delta_i$ of the spikes around the $i^{th}$ through from the Gaussian variance 
\begin{equation} 
\Delta_i^2 = \frac{1}{\beta_{\gamma} A_{\gamma} \omega_i^2}.
\nonumber
\end{equation}
This variance is about six times smaller than the instantaneous period, $6\Delta_i \approx T_i  = 2\pi /\omega_i$.
Hence 
\begin{equation} 
\frac{6}{\sqrt{\beta_{\gamma} A_{\gamma}} \omega_i} \approx \frac{2\pi}{\omega_i }
\nonumber
\end{equation}
which implies that the effective temperature is approximately equal to the amplitude
\begin{equation} 
\frac{1}{\beta_{\gamma}}\approx A_{\gamma,i}.
\nonumber
\end{equation}
Although the ``effective temperature" $\beta_{\gamma}$ may differ between different cell assemblies and different $\gamma$-troughs, 
we consider the simplified case in which a single parameter $\beta$ defines the mean coupling between the $\gamma$-wave's amplitude 
and frequency and the place cell's spike times across the entire place cell network. By normalizing the amplitude $A_{\gamma}$ by the
$\sigma_{\gamma} =\left\langle \sqrt{A^2_{\gamma}-A^2_{0}} \right\rangle_{t}$, $A=A_{\gamma}/\sigma_{\gamma}$, we get the 
scaled parameter $\beta = \beta_{\gamma}\sigma_{\gamma}$, with the characteristic value
\begin{equation} 
\beta = \frac{1}{A}.
\nonumber
\end{equation}
\textbf{Cell types}. The described approach can be applied to both the TroPyr and the RisPyr cells. Mathematically, the ``raising phases of 
$\gamma$" that controls spiking of the RisPyr cells correspond to the vicinities of peaks of the $\gamma$-amplitude's time derivative. 
Hence, the spiking probability of the RisPyr can be constrained by a factor similar to (\ref{Boltz}), involving the derivative of the $\gamma$-amplitude, 
$A'(t)$, which would overridden the $\theta$-precession constraint ($\Lambda_{\theta,c}(\varphi) = 1$) in the vicinity of the $A'(t)$-peaks. 
The analysis of the mixed (RisPyr and TroPyr) ensembles is more complex and requires a discussion \emph{sui generis}.

\textbf{Mathematical methods} required for this study are based on the Persistent Homology Theory (see \cite{PLoS} and 
\cite{Carlsson,Zomorodian}) implemented in ``JPlex" freeware package \cite{JPlex}. 
 
\newpage

\section{References}
\label{section:refs}

 
\newpage
\beginsupplement

\section{Supplementary Figures}
\label{section:SupplFigs}

\begin{figure}[ht]
\includegraphics[scale=0.79]{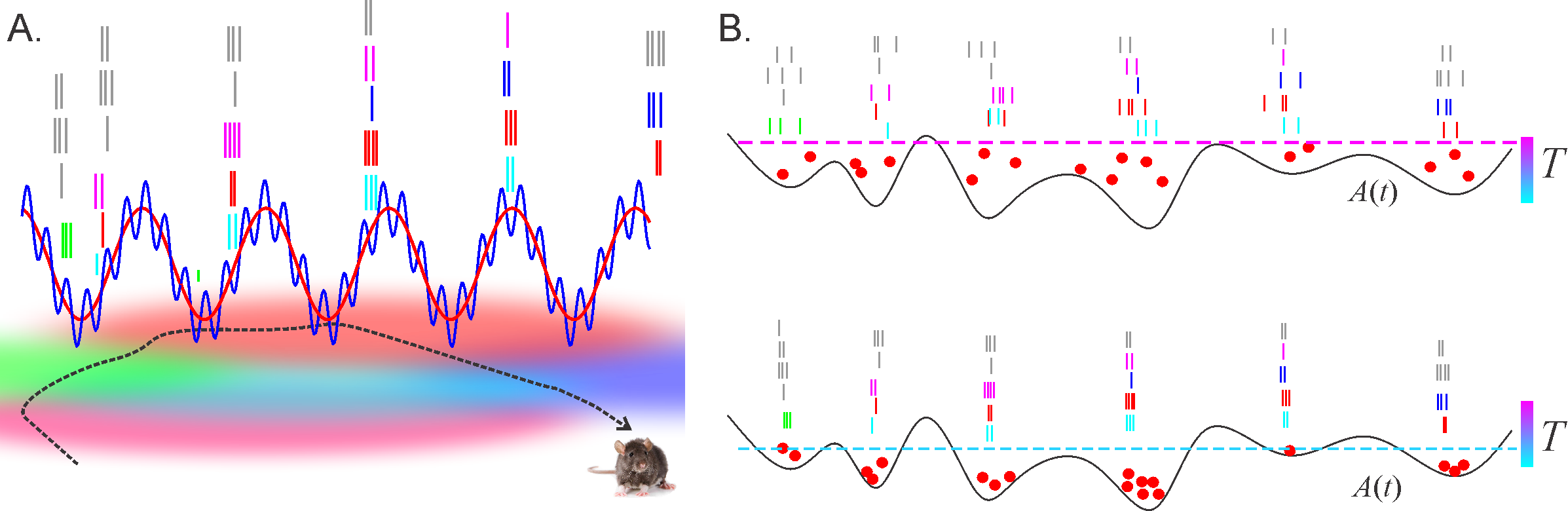}
\caption{\label{SFigure1} \textbf{Brain rhythms modulate place cell spiking activity}. (\textbf{A}). 
Spike times precess with the $\theta$-rhythm (red wave): as the rat progresses through a place field, 
the corresponding place cell discharges at a progressively earlier phase in each new $\theta$-cycle. These 
``preferred" phases of the $\theta$-rhythm correspond to particular $\gamma$-cycles; the blue wave 
shows the net $\theta$ + $\gamma$ amplitude. The synchronized spikes (shown by tickmarks colored 
according to the place fields traversed by the animal's trajectory) cluster over the $\gamma$-troughs, 
yielding dynamical cell assemblies. (\textbf{B}) The spread of spike times around the $\gamma$-troughs, 
by analogy with stochastic particles in a $1D$ potential (black curve). If the temperature is high (dashed 
line, top panel), the particles (red dots) spread diffusely over the potential landscape, and when the 
temperature is low (bottom panel), they are confined at the bottoms of the potential wells. A similar 
effect is produced if the place cells' firing rate is modulating by the Boltzmann factor 
$e^{-\beta_{\gamma}A_{\gamma}(t)}$, where $A_{\gamma}(t)$ is the amplitude of the $\gamma$-wave 
and $\beta_{\gamma}$ represents the inverse temperature. When $\beta_{\gamma}$ is low, the dynamical 
cell assemblies are ``hot" (i.e., more spread in time), and when $\beta_{\gamma}$ is large, the spikes are 
concentrated at the $\gamma$-troughs.} 
\end{figure} 

\begin{figure}[ht]
\includegraphics[scale=1.16]{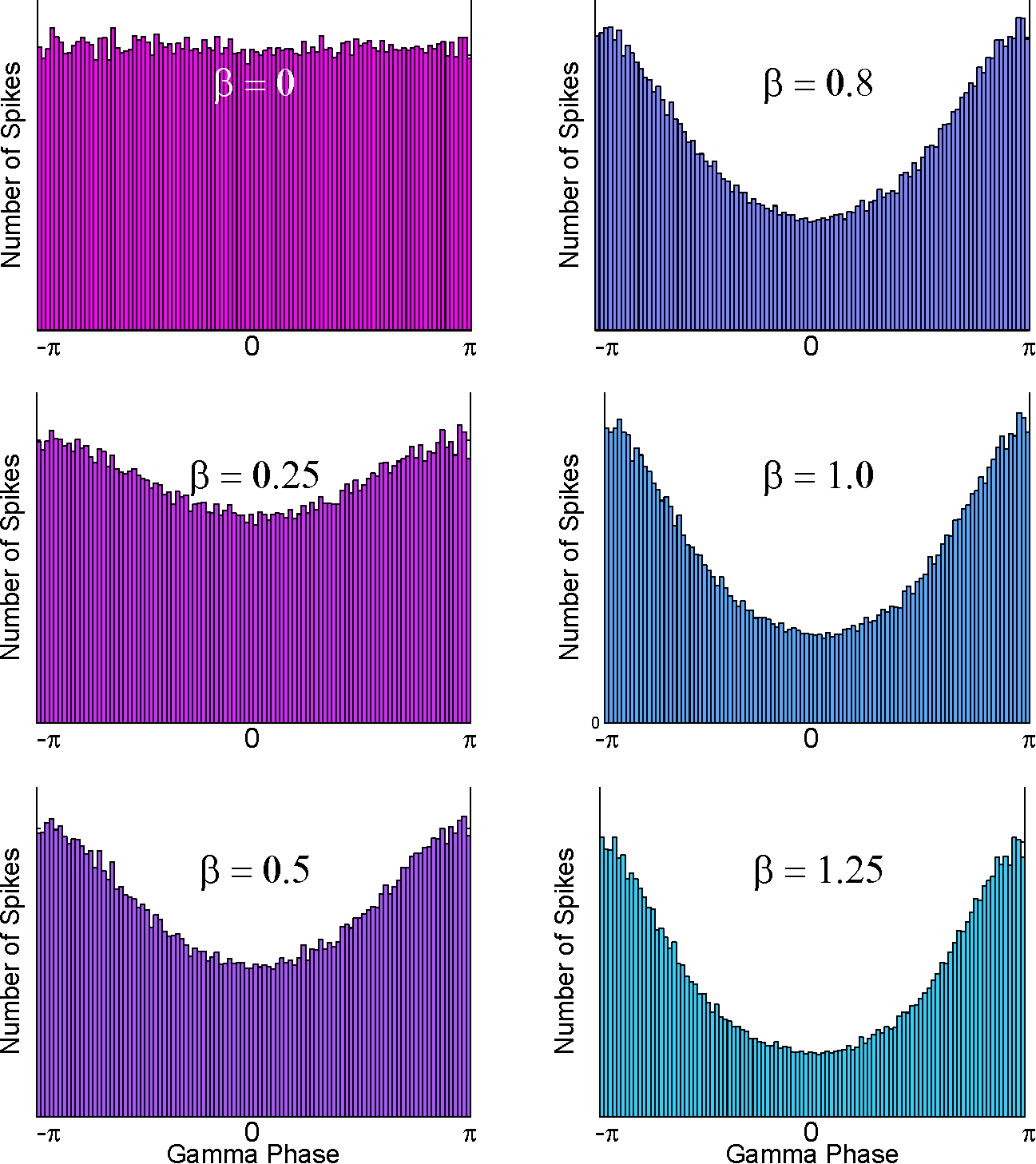}
\caption{\label{SFigure2} \textbf{Histograms of the $\gamma$-phases at the times of place cell spiking}, 
as a function of the inverse effective temperature $\beta$. The cooler the cell assemblies, the more the spikes 
are coupled with the $\gamma$-troughs.} 
\end{figure} 

\begin{figure}[ht]
\includegraphics[scale=0.89]{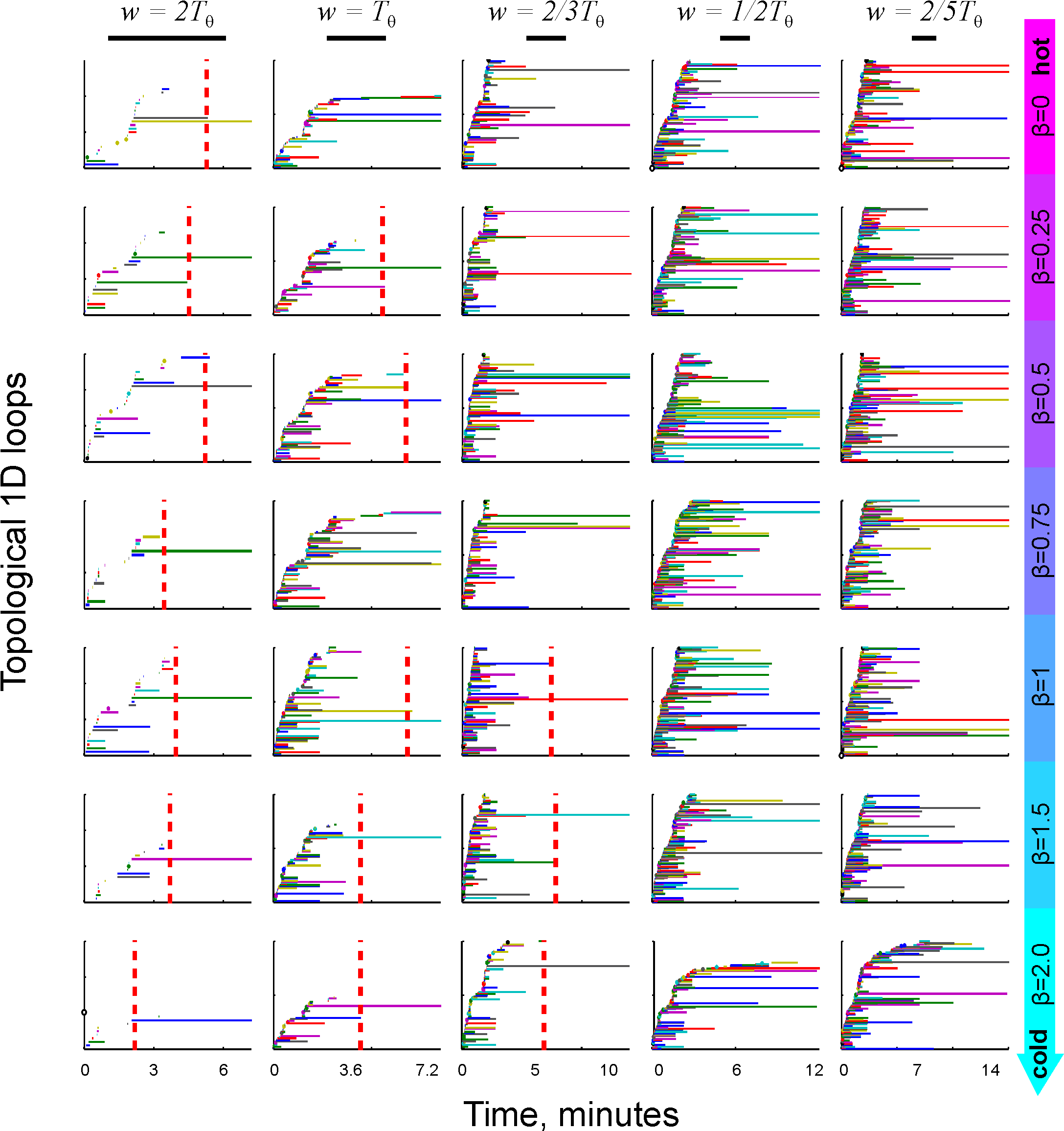}
\caption{\label{SFigure3} \textbf{Freezing out spurious loops}. Timelines of the topological loops in the 
coactivity complex produced in the environment shown in Figure 1 for different integration windows (scale of 
$w$s is shown on top) and for different effective temperatures $1/\beta$ (colorbar on the right). As the width 
of the integration window decreases, the number of spurious topological loops in the coactivity complex increases. 
For large $w$s, spurious loops tend to disappear with learning (the times $T_{min}$ when the correct topological 
structure of $\mathcal{T}_{\sigma}$ emerges are marked by vertical dashed lines). For small $w$s, some of these loops 
persist, indicating that the detected coactivity information is insufficient for eliminating spurious holes in $\mathcal{T}_{\sigma}$. 
However, cooling down the coactivity complex suppresses the proliferation of the spurious loops: at $\beta = 2$ 
(bottom row) the coactivity complex has a correct structure at the integration window $w \approx (2/3)T_{\theta}$.} 
\end{figure} 
 
\begin{figure}[ht] 
\includegraphics[scale=0.89]{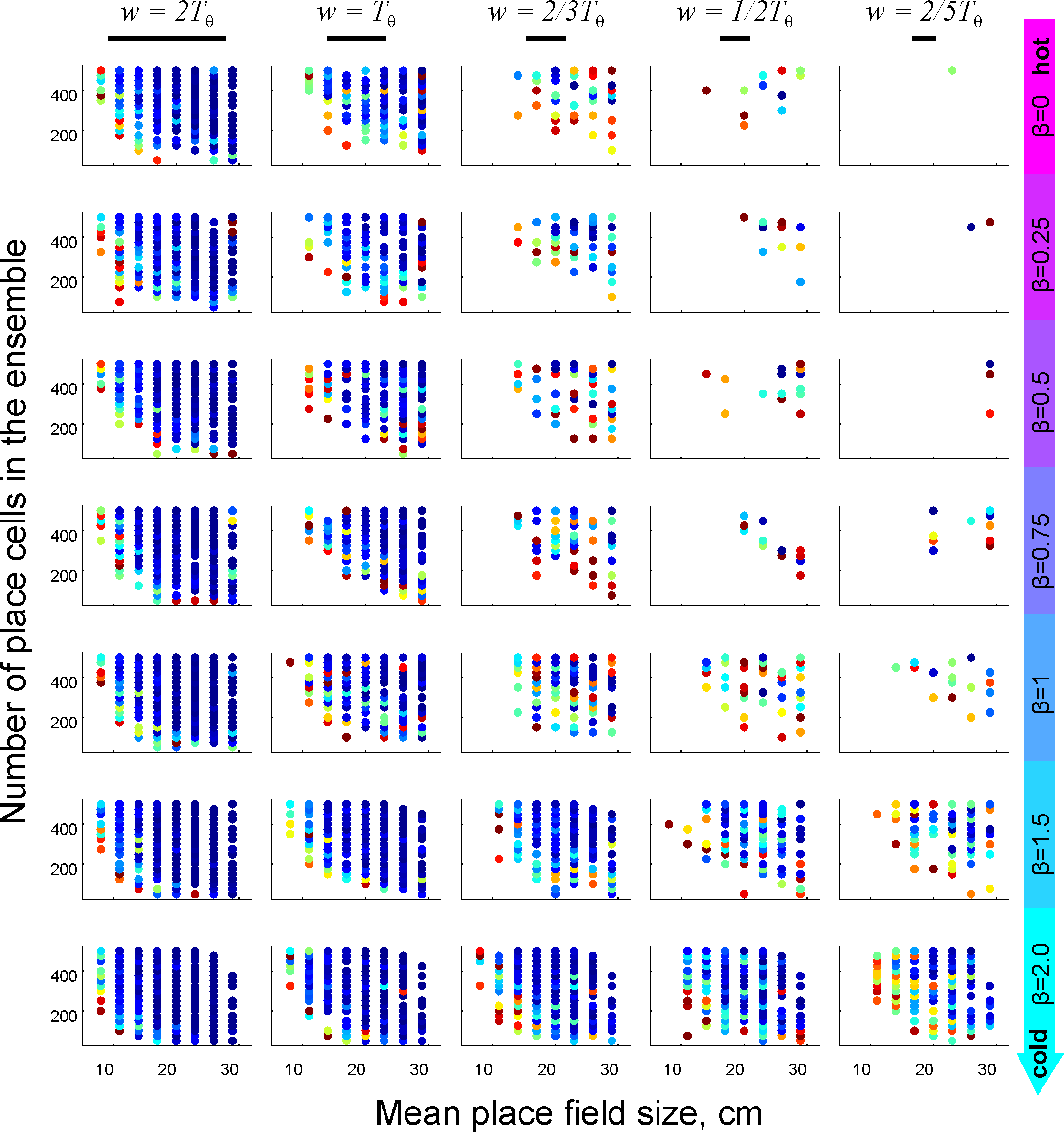}
\caption{\label{SFigure4} \textbf{The effect of $\gamma$-synchronization on spatial learning}. Each panel represents 
the results of simulating 150 neuronal ensembles at different effective temperatures $1/\beta$ (colorbar on the right) 
and different integration times $w$ (scale shown above). Each dot represents a particular ensemble of $N_{c}$ place 
cells with the mean place field size $s$. The maximal firing rates of the simulated neurons are lognormally distributed 
around $f = 25$ Hz (see Methods in \cite{PLoS,Arai}). The color of the dot indicates the average time $T_{min}$ 
required to encode an accurate map of the environment shown on Figure~\ref{Figure2}A, averaged over ten place 
field maps with the same $(s, N)$. If the integration window is large (two left-most columns), $\gamma$-synchronization 
does not produce a strong effect on learning times. As the integration window becomes smaller, cooling the coactivity 
complex increases the scope of successful place cell ensembles. This implies that that $\gamma$-synchronization 
increases the resilience of the hippocampal network in the face of variations of the place-spiking parameters.} 
\end{figure} 

\begin{figure}[ht]
\includegraphics[scale=0.98]{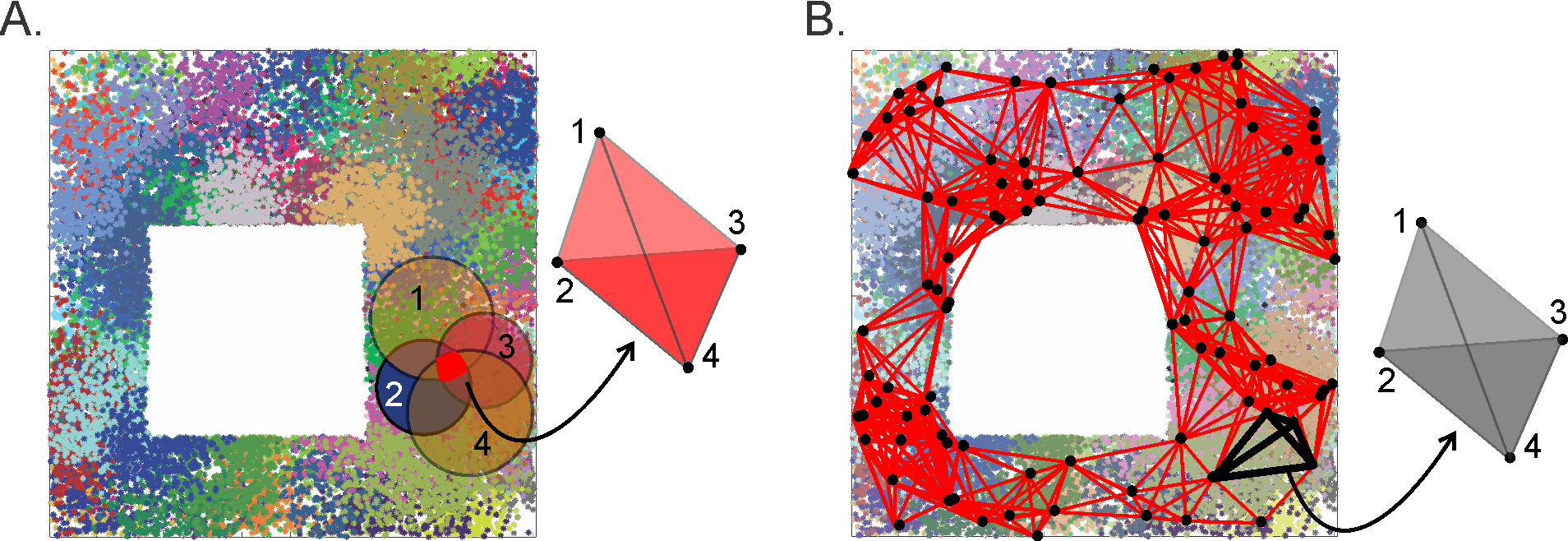}
\caption{\label{SFigure5} \textbf{The coactivity complexes}. (\textbf{A}) Simplexes of the \v{C}ech coactivity 
complex represent simultaneous overlaps of the place fields. (\textbf{B}) The place field connectivity graph $G$: 
dots represent centers of the place fields, red links represent overlaps between the place fields. The fully connected 
subgraphs of $G$ (e.g., the six black links at the bottom of the panel) are the cliques of $G$ representing simplexes of 
the clique coactivity complex $\mathcal{T}_{\varsigma}$.} 
\end{figure} 
 
\begin{figure}[ht]
\includegraphics[scale=0.89]{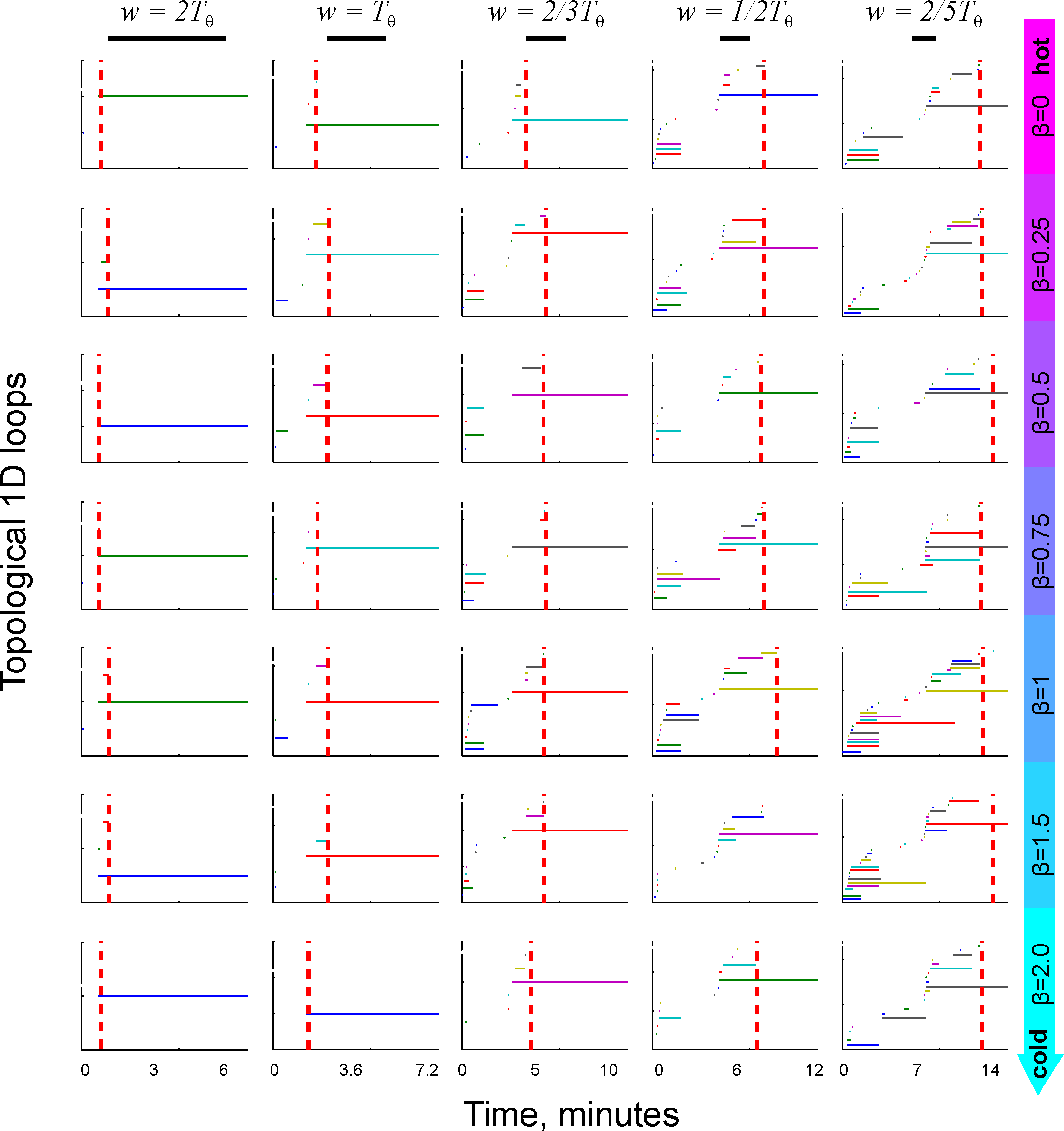}
\caption{\label{SFigure6} \textbf{Freezing out the spurious loops in clique complex}. Timelines of the topological 
loops in the clique coactivity complex produced in the environment shown on Figure~\ref{Figure1}, for different 
integration windows (scale of $w$s is shown on top) and different effective temperatures $1/\beta$ (colorbar on 
the right). 
The learning times $T_{min}$ are marked by red vertical dashed lines. The qualitative dependence of the number of 
topological loops in the coactivity complex on the width of the integration window and the effective temperature 
$1/\beta$ are similar to the ones produced by the coactivity complex. However, the overall numbers of spurious 
topological loops is smaller, and the coactivity complex has a correct structure even at the smallest integration 
window $w \approx (2/5)T_{\theta}$.} 
\end{figure} 

\begin{figure}[ht]
\includegraphics[scale=0.84]{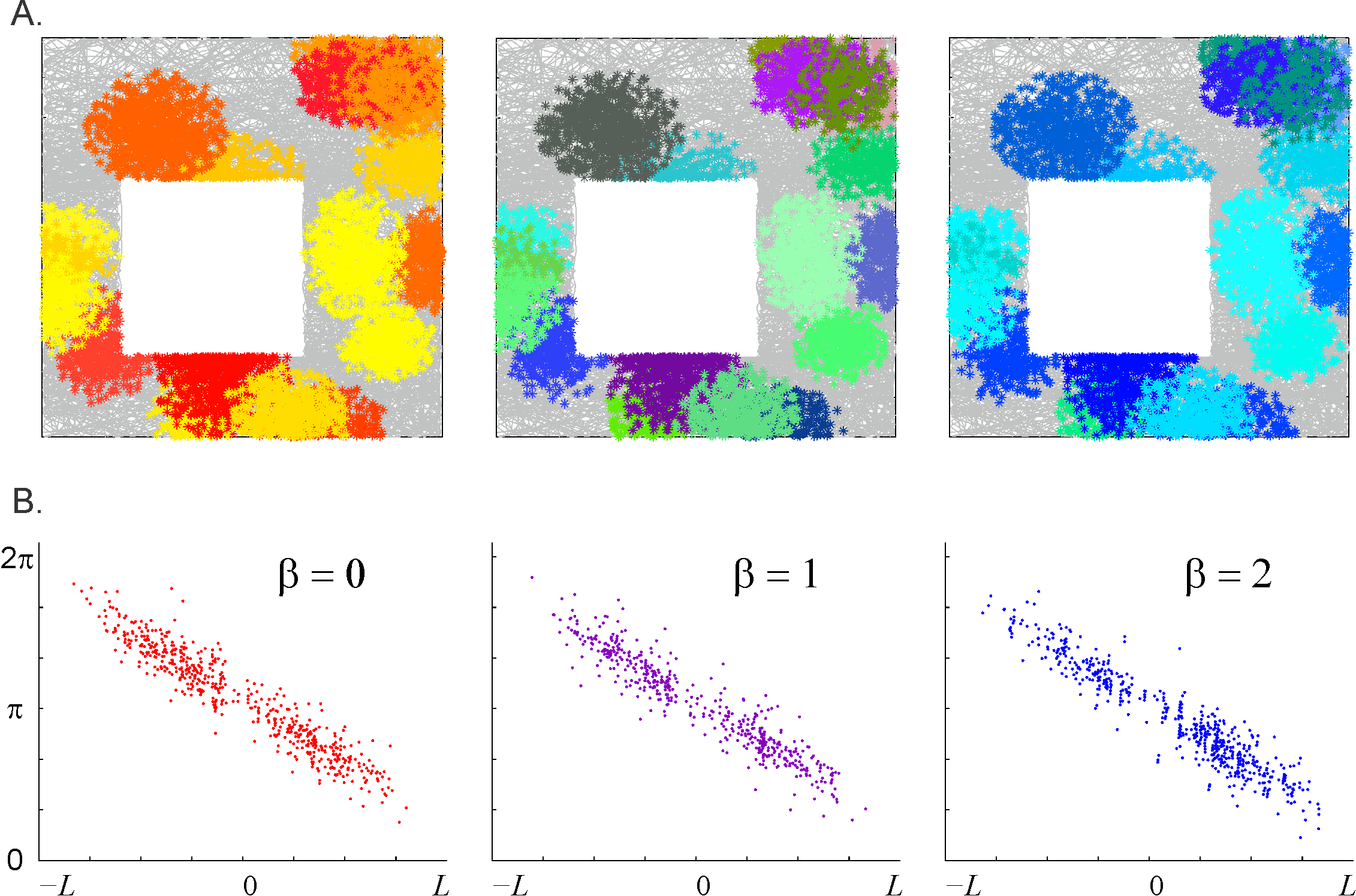}
\caption{\label{SFigure7} \textbf{Simulated place fields and the $\theta$-precession are not affected by the 
gamma modulation}. (\textbf{A}) Place fields shown for $\beta = 0$, $\beta = 1$ and $\beta =2$. 
(\textbf{B}) The $\theta$-phase/position diagram illustrating the $\theta$-precession of a simulated place cell for 
$\beta = 0$, $\beta = 1$ and $\beta = 2$.} 
\end{figure} 


\end{document}